\documentclass[aps,pra,groupedaddress,floats,reprint]{revtex4-2}

\usepackage[english] {babel}
\usepackage{amsfonts,graphicx,soul,natbib,mathrsfs,dsfont}

\usepackage[colorlinks=true]{hyperref}

\usepackage{epstopdf}
\usepackage{xcolor}
\usepackage{soul}
\usepackage{physics}
\usepackage{graphics}
\usepackage{comment}

\renewcommand{\vec}[1]{\textnormal{\boldmath$#1$}}

\usepackage{tikz}
\usepackage{amscd}
\usepackage[inline]{enumitem}
\usepackage[
    left=1.9cm,
    right=1.9cm,
    top=1.5cm,
    bottom=2cm,
    bindingoffset=0cm,
]{geometry}
\usepackage{amsmath}
\usepackage{cancel}
\usepackage{physics}

\usepackage[normalem]{ulem}

\usepackage{dcolumn}
\usepackage{bm}
\begin{document}

\title{Unitary equivalence of twisted quantum states}

\author{N. V. Filina}%
\affiliation{School of Physics and Engineering,
ITMO University, St. Petersburg, Russia 197101}%

\author{S. S. Baturin}%
\email{s.s.baturin@gmail.com}%
\affiliation{School of Physics and Engineering,
ITMO University, St. Petersburg, Russia 197101}%

\date{\today}
\begin{abstract}
We present the time dynamics of twisted quantum states. We find an explicit connection between the well-known stationary Landau state and an evolving twisted state, even when the Hamiltonian accounts for  linear energy dissipation. Utilizing this unitary connection, we analyze nonstationary Landau states and unveil some of their properties. The proposed transformation enables simple evaluation of different operator mean values for the evolving twisted state based on the solution to the classical Ermakov equation and matrix elements calculated on the stationary Landau states. The suggested formalism may significantly simplify analysis and become a convenient tool for further theoretical development on the dissipative evolution of the twisted quantum wave packet.      
            
\end{abstract}

\maketitle

\section{Introduction}
Cylindrical waves are a well-known phenomenon in the field of waveguides and antennas \cite{Vain}. Such waves naturally appear as a solution to the D'Alambert equation in a cylindrical coordinate system as a consequence of the cylindrical symmetry of the problem. 

In \cite{Allen1,Allen1999} Allen and coauthors realized that cylindrical waves could be generated in free space in the optical regime. Moreover, they pointed out that the azimuthal index, $l$, of the cylindrical electromagnetic wave corresponds to the quantized projection of the angular momentum of light. Indeed, cylindrical symmetry implies invariance of the solution to elementary rotations along the symmetry axis. The generator for cylindrical symmetry is the $-i\partial_\phi $ operator. Thus, the solutions of the corresponding problem must be a superposition of the eigenfunctions of this operator, which is known to form a complete set under periodic boundary conditions. Eigenvalues $l$, with $l\in \mathbb{Z}$, enumerate basis functions and correspond to the orbital angular momentum, as, by definition, $\hat{L}_z=-i\partial_\phi$. With the development of the Berry theory, it was pointed out that the orbital angular momentum is just a coefficient in a Berry phase (or geometric phase) defined by \cite{Berry1,Berry2}

\begin{align}
\label{eq:charge}
    l=\lim\limits_{r\to \infty}  \frac{1}{2\pi} \int\limits_{0}^{2\pi}d\phi \frac{d \left(\arg \psi \right)}{d\phi} , 
    \end{align}
where $\psi$ is the particle (photon, electron, proton, etc.) wave function normalized to unity, $\langle \psi|\psi \rangle=1$. On the other hand, for the case of the defined orbital angular momentum (when the wave function factorizes into the radial and angular parts), the formula above could be reinterpreted as
\begin{align}
        l=\lim\limits_{r\to \infty}\frac{1}{4\pi|\psi|^2}\int\limits_{0}^{2\pi} \Im \left[\psi^* \partial_\phi \psi \right] d \phi=\langle \psi| \hat L_z | \psi \rangle, 
\end{align}
unveiling the connection between the strength of the vortex singularity and the eigenvalue of the $\hat{L}_z$ operator. As a result, such waves are often referred to as waves with nontrivial geometric phase, or waves that possess a phase vortex, as well as, waves that carry orbital angular momentum (OAM). 

Numerous applications of light beams with nonzero OAM, or twisted photons, have been widely discussed \cite{ct3,FrankeArnold2008,Mono,UFN}. It has also been shown that electrons with a phase vortex - twisted electrons - could be a new tool for microscopy, materials science, and high-energy physics \cite{ct4,IvanovPubl}. 

While significant understanding was gained on the theoretical side \cite{ct1,JAGA89,JAGA90,JAGA95,Bliokh2007,Bliokh2012,FarEff,NUF21,Karlovets_paraxial1,Karlovets2021}, there are still several challenges in the complete description of the evolution of the twisted particles. For instance, radiation and the possible consequent loss of angular momentum is an open question, as it relates to the acceleration of twisted massive charged particles to relativistic energies. 

In the present study, inspired by the historical connection of the twisted states to geometry, we apply a geometric transformation of the classical particle phase trajectory \cite{Arnold} and build a corresponding transformation for Schr\"{o}dinger's equation \cite{QAT1,QAT2,QAT3}. It is shown that the simple idea of vector field rectification, discussed by Arnold in his famous book \cite{Arnold}, has far-reaching consequences \cite{QAT2,QAT3}. In the present study we take the course of Refs.\cite{QAT1,QAT2,QAT3} called QAT (the Quantum Arnold Transformation) as we find it the most intuitive. A comprehensive study of the time-dependant quantum harmonic oscillator can be found in Ref.\cite{Manko}.   

The benefit of QAT approach is twofold. First, it allows mapping any state that satisfies Schr\"{o}dinger's equation for the case of a temporally constant (or constant in the longitudinal coordinate $z$ with the paraxial approximation) magnetic field (refractive index for the case of the photons) to the case of time-varying potential, including a system with linear dissipation within the Cardirola-Kanai (CK) model \cite{Cardirola,Kanai}. The latter, in turn, enables accounting of media absorption for the case of neutral particles, and the radiation friction in the dipole approximation for the case of charged particles. 
Second, it allows calculation of the mean values of the observable operators, based on the corresponding matrix elements from the solution of the system with stationary (time-independent) potential.   

Throughout the paper we use the natural unit system with $\hbar = 1$, $c = 1$, and assume that $e<0$ is the electron charge.

\section{QAT formalism and 2D Ermakov operator}
We analyze the transverse part of the nonrelativistic Schr\"{o}dinger equation for a massive charged particle in a magnetic field that has the form 

\begin{align}
\label{eq:mainSch}
    i\partial_t \psi=\hat H_\perp \psi,~~~\hat{H}_\perp=\frac{\left[\hat{\vec{p}}_\perp-e \vec{A}(t) \right]^2}{2 m}. 
\end{align}
Here $\vec{A}^T=\left\{-y B(t)/2,x B(t)/2\right\}$ is the transverse part of the vector potential, and $B(t)$ is the modulus of the magnetic field directed along the $z$-axis. 

Equation~\eqref{eq:mainSch} is the basic equation for  describing the evolution of the transverse part of the twisted particle wave function \cite{SilenkoG,Bliokh2007}, either in the paraxial approximation for both photons in free space and fibers \cite{N2opt} (with the proper change in the parameters \cite{Bliokh2007}), and for relativistic electrons (after factoring out the longitudinal part and substitution of $t\to z$) \cite{Silenko2021}. Using a point particle approximation \cite{Baturin}, Eq.\eqref{eq:mainSch} describes the transverse dynamics of a non-relativistic electron traversing a set of solenoid lenses. In all cases, the transverse part of the wave function $\psi$ defines the twisted structure of the corresponding state. In the present paper for definiteness, we consider electrons in a magnetic field. The same approach is valid for other types of particles where Eq.\eqref{eq:mainSch} describes transverse dynamics of the wavefunction. 

If we account for the linear friction that may arise, for instance, from radiation friction, then Schr\"{o}dinger's equation generalizes to the CK model by a canonical transformation \cite{Cardirola,Kanai} 
\begin{align}
\label{eq:CKH}
    i\partial_t \psi&= \\ \nonumber&\left[\frac{w(t)\left[\hat{p}_x^2+\hat{p}_y^2\right]}{2m}+\frac{m\omega^2(t) (\hat{x}^2+\hat{y}^2)}{2 w(t)}+\omega(t) \hat{L}_z\right]\psi. 
\end{align}
Where $w(t)$ is the dissipation factor of the form $w(t)=\exp\left[-\int \gamma(t) dt\right]$ with $\gamma(t)$ being the classical friction coefficient. We note that Eq.\eqref{eq:CKH} is the sum of the Hamiltonians of two one-dimensional quantum oscillators coupled through the $\omega \hat{L}_z$ term. 

Here 
\begin{align}
 \hat{L}_z=\hat{x}\hat{p}_y-\hat{y}\hat{p}_x    
\end{align}
is the operator of the angular momentum $z$-projection, and
\begin{align}
\label{eq:frt}
    \omega(t)=\frac{|e|B(t)}{2m}.
\end{align}

A CK system is equivalent to a master equation for the density matrix without fluctuation, assuming there is no stochastic process. However, a full equivalence of the CK model to the standard Lidnblad master equation \cite{Lindblad,Lindblad2} could be achieved once the proper stochastic potential is incorporated into Eq.\eqref{eq:CKH} \cite{Lugiato}. For the present consideration, we limit ourselves to a simplified CK model but note that the analysis we present is extendable to the case of momentum and coordinate diffusion.

Now, we consider the quantum Arnold transformation (QAT) discussed in Refs.\cite{QAT1,QAT2,QAT3}. For a one-dimensional system, the transformation is an interchange of the coordinates, time, multiplication of the wave function by a phase factor, and a normalization factor. 

Following \cite{QAT1} we consider the QAT operator that establishes a mapping between the Hilbert space $\mathcal{H}_t$ of solutions $\psi(x,t)$ of the one-dimensional time-dependent Schr\"{o}dinger equation at time $t$, on the Hilbert space $\mathcal{H}_\tau$ of solutions $\varphi(\kappa,\tau)$ of the time-dependent Schr\"{o}dinger's equation for the Galilean free particle at a time $\tau$. The explicit form for the 1D QAT is (see Appendix \ref{QATcl} and Appendix \ref{QATq})
\begin{equation}
\label{eq:QAT}
\hat{\mathcal{Q}}:\begin{cases}
    \kappa = \frac{x}{u_2}, \\
    \tau = -\frac{u_1}{u_2}, \\
    \varphi(\kappa, \tau) = \psi\left(x, t\right) \sqrt{u_2} \exp\left[{-\frac{i}{2} \frac{m}{w} \frac{\dot{u}_2}{u_2}x^2}\right].
\end{cases}
\end{equation}
In the equality for the wave function, $\varphi$, the arguments $x$ and $t$ on the right, as well as the argument $t$ of the functions $u_2$ and $w$, are understood as functions of $\kappa$ and $\tau$ ($x=x(\kappa,\tau)$; $t=t(\tau)$). The dot above $u_2$ denotes the total derivative by $t$. 
The functions $u_1$ and $u_2$ are general solutions to the classical Euler-Lagrange equation of the considered quantum oscillator
\begin{align}
\label{eq:EL}
     \ddot{u}_{1,2}+\gamma(t) \dot{u}_{1,2}+\omega^2(t) u_{1,2}=0 
\end{align}
and 
\begin{align}
\label{eq:WR}
    w(t)=u_1 \dot{u}_2-u_2 \dot{u}_1=\exp[-\int\gamma dt] 
\end{align}
is the Wronskian built on these solutions. The initial conditions for Eq.\eqref{eq:EL} are chosen such that linear independence of $u_1$ and $u_2$ is guarantied and Eq.\eqref{eq:WR} holds.

The first part of the transformation is simply the coordinate and time substitution that transforms part of a curved classical phase trajectory in the extended phase space to a straight line \cite{Arnold,QAT1}. The coordinate transformation induces a gauge transformation for the vector potential. Indeed, the classical momentum could be expressed as
\begin{align}
    p_\tau = m \frac{d \kappa}{d\tau}= \frac{dt}{d\tau}  \frac{p_x}{u_2}-m \frac{\dot{u_2}}{u_2^2} \frac{d t}{d \tau} x,
\end{align}
and consequently, the gauge potential \cite{davidov} is 
\begin{align}
    G=-m \frac{\dot{u_2}}{u_2^2} \frac{d t}{d \tau} x=-m \frac{\dot{u_2}}{w}x.
\end{align}
Therefore, the phase of the wave-function induced by the transformation of time and coordinate is 
\begin{align}
    &\theta_\kappa=-m \frac{\dot{u_2}}{w} \int\limits_{0}^x \tilde{x} d\kappa=\\ \nonumber&-\frac{m}{w} \frac{\dot{u_2}}{u_2} \int\limits_{0}^x \tilde{x} d \tilde{x}=-\frac{m}{2w} \frac{\dot{u_2}}{u_2} x^2.
\end{align}
and we recover the phase multiplier of the wave-function in Eq.\eqref{eq:QAT}.    

Schr\"{o}dinger's equation is invariant under $\hat{\mathcal{Q}}$, consequently 
\begin{align}
    i \frac{\partial \varphi}{\partial \tau} &= - \frac{1}{2 m} \frac{\partial^2 \varphi}{\partial \kappa^2}, \nonumber \\ 
    &\hat{\mathcal{Q}}^{-1}~~\Downarrow \\
     i \frac{\partial \psi}{\partial t} &= -\frac{w}{2 m} \frac{\partial^2 \psi}{\partial x^2} + \frac{m \omega^2 x^2\psi}{2 w}. \nonumber
\end{align}
The latter could be checked, for example, by direct substitution (see Appendix \ref{QATq} for the details).
The transformation holds up to a common multiplier $\frac{u^2_2 \sqrt{u_2}}{w}$ in the second equation and is local in time, as it is valid for $u_2\neq 0$. 

We reiterate, $\hat{\mathcal{Q}}^{-1}$ and $\hat{\mathcal{Q}}$ are operations that transforms the free 1D Schr\"{o}dinger's equation to the equation of 1D CK system and back.

For a given 1D system operator $\hat{\mathcal{Q}}$ is unique. This follows from the fact that $\hat{\mathcal{Q}}^*\hat{\mathcal{Q}}=\hat{I}$, i.e. $\hat{\mathcal{Q}}$ is a unitary operator. The latter follows from the fact that the transformation $\hat{\mathcal{Q}}$ preserves the norm i.e. $||\phi_x||_{\mathcal{H}_\tau}=||\psi_x||_{\mathcal{H}_t}$. Indeed, the Jacobian determinant of the transformation for the fixed moment in time, as it follows from the coordinate substitution, is just $1/u_2$, i.e. $d\kappa=d x/u_2$ 
while the square of the amplitude of the wave function calculates to $|\phi|^2=|\psi|^2 u_2$. Consequently $\int |\phi|^2 d\kappa = \int |\psi|^2 dx$.  

\begin{figure}[t]
\center{\includegraphics[width=0.6\linewidth]{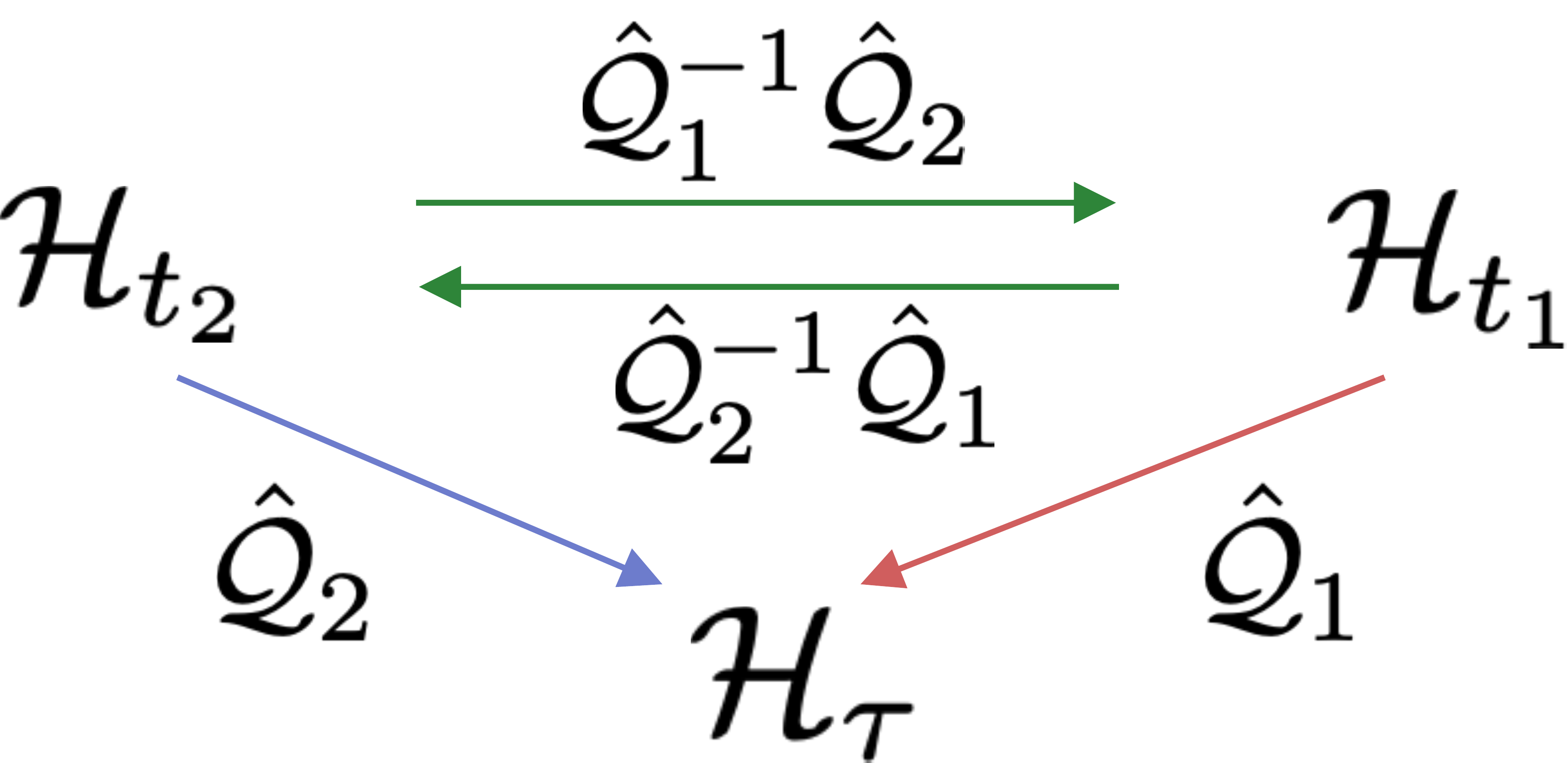}}
\caption{Schematic diagram of the combined QAT mapping \cite{QAT2,QAT3}.}
\label{Fig:1}
\end{figure}

Next, we consider two 1D oscillators that differ by their dissipation factors, $w_{1,2}$, and oscillator frequencies, $\omega_{1,2}$. Using the QAT formalism, both oscillators could be mapped to a free particle (Hilbert space,  $\mathcal{H}_\tau$). As a result, the QAT combination (as shown in the diagram Fig.\ref{Fig:1}) maps the Hilbert space of one oscillator, $\mathcal{H}_{t_1}$ , to the Hilbert space of another, $\mathcal{H}_{t_2}$.

From Eq.\eqref{eq:QAT}, we derive \cite{QAT3} (see Appendix \ref{Ermakov1D})
\begin{equation}
\label{eq:EP}
\hat{\mathcal{E}}_{1 \mapsto 2}:\begin{cases}
    x_1 = \frac{x_2}{b(t_2)}, \\
    w_1(t_1)dt_1 = \frac{w_2(t_2)}{b^2(t_2)}dt_2, \\
    \psi_2(x_2, t_2) = \psi_1\left(x_1, t_1 \right) \frac{\exp\left[{\frac{i}{2} \frac{m}{w_2} \frac{\dot{b}}{b}x_2^2}\right]}{\sqrt{b}},
\end{cases}
\end{equation}
where we have introduced the Ermakov mapping operator $\hat{\mathcal{E}}_{1\mapsto 2}\equiv \hat{\mathcal{Q}_2}^{-1}\hat{\mathcal{Q}_1}$, and $b(t_2)$ - a scaling parameter that satisfies the Ermakov-Pinney equation \cite{Ermakov,Pinney} with damping
\begin{align}
\label{eq:EPeq}
    \ddot{b} + \gamma_2(t_2) \dot{b} + \omega^2_2(t_2) b = \frac{w^2_2(t_2)}{w^2_1(t_1)} \frac{\omega^2_1(t_1)}{b^3}.
\end{align}
In general, $\omega_1$ and $w_1$ depend on $t_1$, expressed through $t_2$. The dots above $b$ in Eq.\eqref{eq:EP} and Eq.\eqref{eq:EPeq} indicate total derivatives by $t_2$. 
A direct check of this transformation, as well as the derivation of the Eq.\eqref{eq:EPeq}, is provided in the Appendix \ref{Ermakov1D}.

The Eramkov operator $\hat{\mathcal{E}}_{1\mapsto2}$, in full analogy with the QAT operator $\hat{\mathcal{Q}}$, maps the Hilbert space $\mathcal{H}_{t_1}$ of solutions $\psi_1(x_1,t_1)$ of the 1D time-dependent Schr\"{o}dinger equation of the first oscillator at time $t_1$, on the Hilbert space $\mathcal{H}_{t_2}$ of solutions $\psi_2(x_2,t_2)$ of the 1D time-dependent Schr\"{o}dinger's equation of the second oscillator at time $t_2$.

To apply the Ermakov operator to the Eq.\eqref{eq:CKH}, one needs to extend it to the 2D case. When the magnetic field is uniform and is directed strictly along the $z$-axis, frequencies of both quantum oscillators in $x$ and $y$ are identical. Consequently, we construct a 2D Ermakov operator as follows
\begin{equation}
\label{eq:EP2d}
\hat{\mathcal{E}}^{2D}_{1 \mapsto 2}:\begin{cases}
    x_1 = \frac{x_2}{b(t_2)}, \\
    y_1 = \frac{y_2}{b(t_2)}, \\
    w_1(t_1)dt_1 = \frac{w_2(t_2)}{b^2(t_2)}dt_2, \\
    \psi_2(x_2,y_2, t_2) = \\~~~~~ \frac{1}{b} \psi_1\left(x_1,y_1, t_1 \right) \chi(x_2,y_2,t_2).
\end{cases}
\end{equation}
Here $\chi(x_2,y_2,t_2)$ is the phase factor given by
\begin{align}
\label{eq:EPph}
    \chi(x_2,y_2,t_2)=&\exp\left[{\frac{i}{2} \frac{m}{w_2} \frac{\dot{b}}{b}(x_2^2+y_2^2)}\right]\times \\ \nonumber &\exp\left[-il\int \omega_2 dt_2+il\int \omega_1dt_1\right].
\end{align}
with $l$ being the eigenvalue of the $\hat{L}_z$ operator. Here, as before, $b(t_2)$ satisfies Eq.\eqref{eq:EPeq}.

Due to the symmetry of the transformation, the ratio, $y_1/x_1$, is conserved and polar angles 

\begin{align}
\label{eq:ang}
 \phi_1=\arctan(y_1/x_1)=\arctan(y_2/x_2)=\phi_2   
\end{align}
 have a one-to-one correspondence. This fact immediately leads to the conservation of the orbital angular momentum $l$ of the twisted state under the transformation given by the Ermakov operator, i.e. $\langle \psi_2|\hat{L}_{z_2}|\psi_2\rangle=\langle \psi_1|\hat{L}_{z_1}|\psi_1\rangle$. 

For the case of twisted states, it is convenient to represent the Ermakov operator in polar coordinates where the 
transformation becomes essentially 1D, with the only exception of the multiplier in the wave function. Switching from $x,y$ to $\rho,\phi$ and using $\rho=\sqrt{x^2+y^2}$ and Eq.\eqref{eq:ang}, we arrive at the expression for the Ermakov operator in  polar coordinates
\begin{equation}
\label{eq:EP2dc}
\hat{\mathcal{E}}^{2D}_{1 \mapsto 2}:\begin{cases}
    \rho_1 = \frac{\rho_2}{b(t_2)}, \\
    w_1(t_1)dt_1 = \frac{w_2(t_2)}{b^2(t_2)}dt_2, \\
    \psi_2(\rho_2,\phi, t_2) = \frac{1}{b} \psi_1\left(\rho_1,\phi, t_1 \right) \chi(\rho_2,t_2).
\end{cases}
\end{equation}
The operators in Eq.\eqref{eq:EP2d} and Eq.\eqref{eq:EP2dc} establish a unique unitary transformation of the states of one two-dimensional quantum harmonic oscillator into another (see Appendix \ref{Erakov2D}), and provide a one-to-one correspondence between twisted states of different types (stationary and dynamic). Thus, we conclude that all dynamic twisted states, including free states, non-stationary states, and states that may appear in different time-dependent fields, are unitarily equivalent to stationary states. This observation opens a convenient way to analyze the properties of dynamic states based on their stationary counterparts.

\section{Applications}
In this section, we consider several examples that illustrate the power of the Ermakov mapping formalism. 

We start from a well-known Landau model and recover in a simple fashion non-stationary Landau states discussed in Ref.\cite{Silenko2021}. We provide formulas for the energy and the mean square radius of the non-stationary Landau state and reveal nontrivial features of these states.

Next, we consider a representative numerical example that provides further insight into the evolution of the twisted electron in the axisymmetric magnetic field. We point out that the evolution of the twisted electron is essentially classical. To complete the analogy with the classical electron in the magnetic field we introduce a quantum emittance operator that is a quantum analog of the corresponding classical geometric emittance.

We conclude by providing a formula for the current operator for the twisted charged massive particle in the general time-dependent magnetic field that can be readily used to analyze the interaction of these particles with the electromagnetic vacuum and to study various scattering processes.   

We note that the simplified model Eq.\eqref{eq:CKH} and its extension to the mean field interaction with the electromagnetic bath have found a variety of applications in nonlinear optics \cite{LLE,LasP1,LasP2,LasP3,LasP4,LasP5,LUG,OptSw}. One of the widely known extensions is the LLE model \cite{LLE,LLEls} for spontaneous spatial pattern formation \cite{LasP1}. We point out that it is most likely that the formalism and ideas developed in the field of lasers and nonlinear optics can be, with some modifications, applied to the description of the massive charged vortex particles.  

\subsection{Non-stationary Landau states}

We chose Landau model \cite{Landau,Bliokh2012} as a reference system with $w_1=1$ and $\omega_1=\omega_0=const$:  
\begin{align}
\label{eq:DAU}
    i\partial_t \psi_1=\left[\frac{\left[\hat{p}_x^2+\hat{p}_y^2\right]}{2m}+\frac{m\omega^2_0 (\hat{x}^2+\hat{y}^2)}{2}+\omega_0 \hat{L}_z\right]\psi_1. 
\end{align}
The solution is a stationary Landau state given by
\begin{align}
    \psi_1(\rho,\phi,t) =N \left(\frac{\rho}{\rho_H}\right)^{|l|}&\mathcal{L}_{n}^{|l|}\left[\frac{2\rho^2}{\rho_H^2}\right] \times \nonumber \\&\exp\left[-\frac{\rho^2}{\rho_H^2} + i l \phi-i \varepsilon_\perp t\right],
    \label{eq:Landau}
\end{align} 
where $\mathcal{L}_{n}^{|l|}$ is the generalized Laguerre polynomial, $n$ is the radial quantum number and $l$ is the orbital angular momentum, 
\begin{align}
\rho_H=\sqrt{\frac{4}{|e|B_0}}=\sqrt{\frac{2}{m \omega_0}}
\label{eq:rhom}
\end{align}
is the characteristic radius of the orbit and 
\begin{align}
    \varepsilon_{\perp}=\omega_0\left(2n+|l|+l + 1 \right) 
    \label{eq:LandauE}
\end{align}
is the transverse part of the energy. 

Now, we consider a mapping of the Landau system onto itself, and Eq.\eqref{eq:EPeq} reduces to 
\begin{align}
\label{eq:EPeqDau}
    \ddot{b} + \omega_0^2 b = \frac{\omega^2_0}{b^3}.
\end{align}
With Eq.\eqref{eq:EP2dc} and Eq.\eqref{eq:EPph} we immediately recover the non-stationary Landau state discussed in detail in Ref.\cite{Silenko2021}

\begin{align}
    &\psi_2(\rho,\phi,t) = \frac{N}{b} \left(\frac{\rho}{b\rho_H}\right)^{|l|}\mathcal{L}_{n}^{|l|}\left[\frac{2\rho^2}{b^2\rho_H^2}\right]\exp\left[-\frac{\rho^2}{b^2\rho_H^2} \right] \times \nonumber \\&\exp\left[-il\omega_0t + i l \phi+{\frac{im}{2}\frac{\dot{b}}{b}\rho^2}-i (\varepsilon_\perp-\omega_0 l) \int \frac{dt}{b^2}\right].
\end{align}
The Ermakov operator is a convenient tool for the analysis of different matrix elements based on the known mean values of the stationary Landau state. For instance, $\langle \rho_2^2 \rangle_2$ is simply 
\begin{align}
\label{eq:r2}
  \langle \rho_2^2 \rangle_2=b^2 \langle \rho_1^2 \rangle_1,   
\end{align}
where 
\begin{align}
\label{eq:rhost}
    \langle \rho_1^2 \rangle_1=\frac{1}{m \omega_0}\left(2n+|l|+1\right).
\end{align}
Indeed, with Eq.\eqref{eq:EP2dc} and accounting for the fact that $dx_2dy_2=b^2dx_1dy_1$, we have
\begin{align}
   \langle \rho_2^2 \rangle_2=\int \psi_2^* \rho_2^2 \psi_2 dx_2dy_2=b^2 \int \psi_1^* \rho_1^2 \psi_1 dx_1dy_1. 
\end{align}
Under the transformation of Eq.\eqref{eq:EP2dc}, the action of the momentum operator transforms to 
\begin{align}
\label{eq:mom_tr}
    \vec{\hat{p}_2} \psi_2 = \chi(\rho_2,t_2)\left[\frac{1}{b^2}\vec{\hat{p}_1} \psi_1 + \frac{m}{w_2} \frac{\dot{b}}{b} \vec{\hat{r}_1} \psi_1 \right],
\end{align}
and, consequently, the mean value of the momentum for a general non-stationary twisted state could be expressed in terms of the mean values of the stationary Landau states  
\begin{align}
    \langle \vec{\hat{p}_2} \rangle_2 = \frac{\langle \vec{\hat{p}_1} \rangle_1}{b} + \frac{m}{w_2} \dot{b} \langle \vec{\hat{r}_1} \rangle_1 ,
\end{align}
with $b$ being a solution to Eq.\eqref{eq:EPeq}, and $\omega_1=\omega_0$ and $w_1=1$. For the non-stationary Landau state one must set $w_2=1$ in the formula above and use Eq.\eqref{eq:EPeqDau} for $b$.  

A more interesting and insightful example is the transformation of the mean energy.
To evaluate the time derivative, we again utilize Eq.\eqref{eq:EP2dc} and get 
\begin{align}
    &\langle n'|i\frac{\partial }{\partial t_2}|n \rangle_2=\frac{\dot{b}}{b}\langle n'|\vec{\hat{p}_1} \vec{\hat{r}_1}|n\rangle_1+\nonumber \\&\frac{1}{b^2}\left[\langle n'|i\frac{\partial}{\partial t_1}|n \rangle_1 -l\omega_0\delta_{n',n}\right]+l\omega_0\delta_{n',n}+\nonumber\\&\frac{m}{2}\left[\omega_0^2 b^2+\dot{b}^2 -\frac{\omega_0^2}{b^2} \right] \langle n'| \hat{r}_1^2|n\rangle_1. 
\end{align}

In the expression above, the bra and ket vectors are not necessarily the same. Strikingly, along with the diagonal term (the mean energy of the non-stationary Landau state $\varepsilon_2\equiv\langle n|i\frac{\partial \psi_2}{\partial t_2}|n \rangle_2$) that reads
\begin{align}
\label{eq:menrNL}
    \varepsilon_2 = \frac{\omega_0(2n+|l|+1)}{2} \left[b^2 + \frac{\dot{b}^2}{\omega_0^2} + \frac{1}{b^2} \right]+ l \omega_0,
\end{align}
we retrieve a non-diagonal term that reflects the mixing of states with different radial indices. This result is in full agreement with that derived by H.R. Lewis in Ref.\cite{Lewis}, however, it comes almost at no effort. Mixing terms in the Hamiltonian matrix results in nonzero probabilities for the corresponding transitions and consequent radiation processes. 

The mean energy of the non-stationary Landau state itself, Eq.\eqref{eq:menrNL}, has an interesting structure as well. Although the first term in Eq.\eqref{eq:menrNL} explicitly depends on time, it is constant as the factor
\begin{align}
\label{eq:fact}
    \frac{1}{2}\left[b^2+\frac{\dot{b}^2}{\omega_0^2} +\frac{1}{b^2} \right]\geq 1
\end{align}
is the first integral of Eq.\eqref{eq:EPeqDau}. 
From the inequality of Eq.\eqref{eq:fact}, it immediately follows that the mean energy of the non-stationary Landau state is always greater than the energy of the stationary Landau state. Along with the off-diagonal terms in the Hamiltonian matrix, and non-vanishing time-dependant quadrupole moment proportional to $b^2$ according to the Eq.\eqref{eq:r2}, this indicates that non-stationary Landau states should be far less stable than the stationary Landau states.

\subsection{Propagation of the twisted electron through a set of solenoids}

\begin{figure}[t]
\center{\includegraphics[width=1\linewidth]{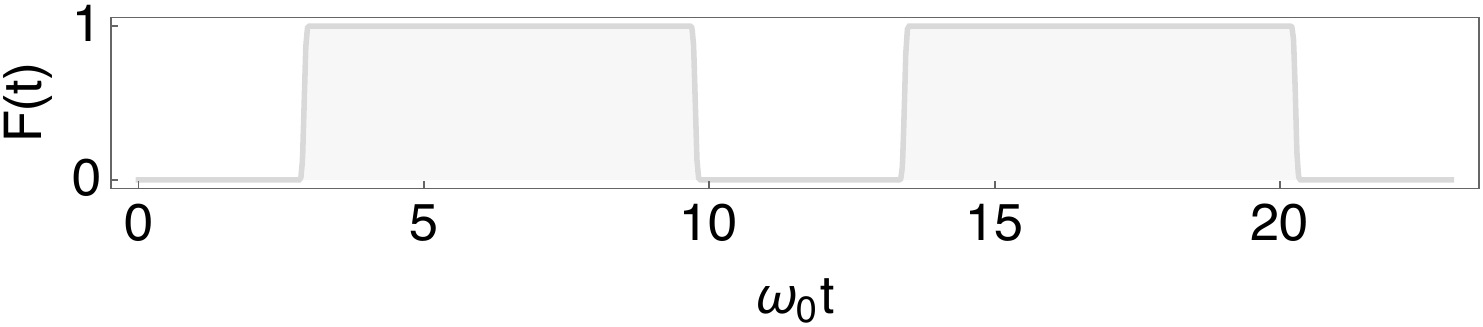}}
\center{\includegraphics[width=1\linewidth]{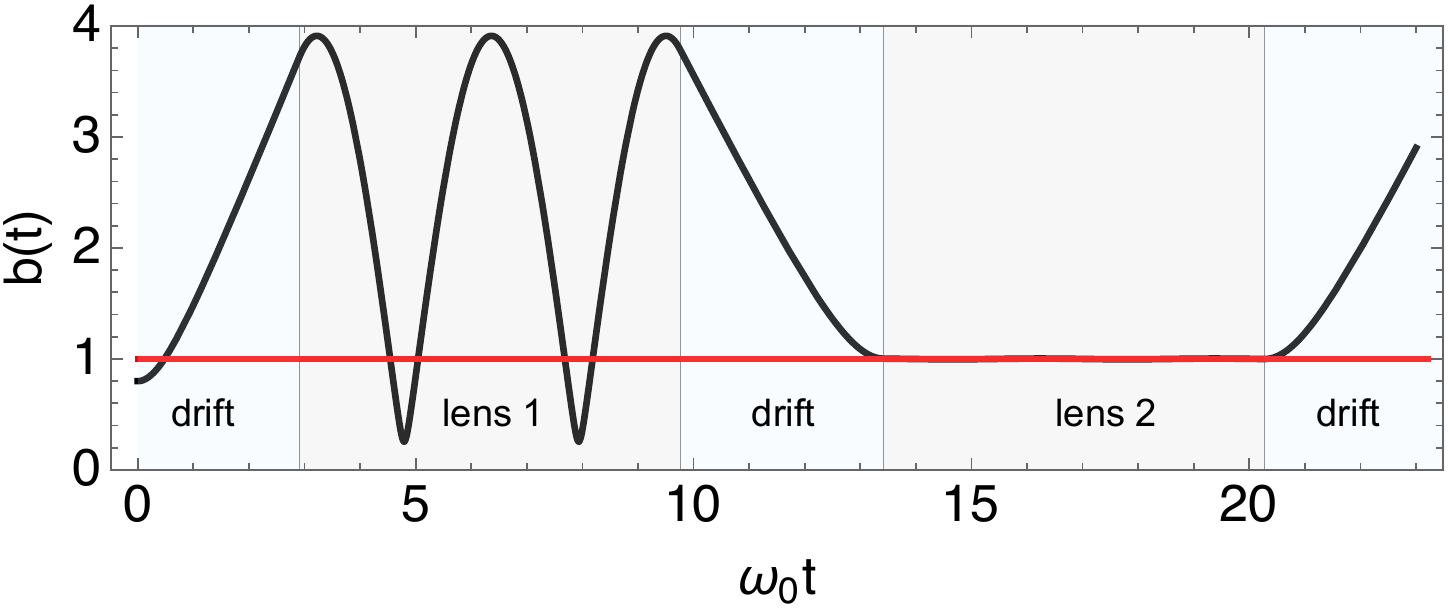}}
\center{\includegraphics[width=1\linewidth]{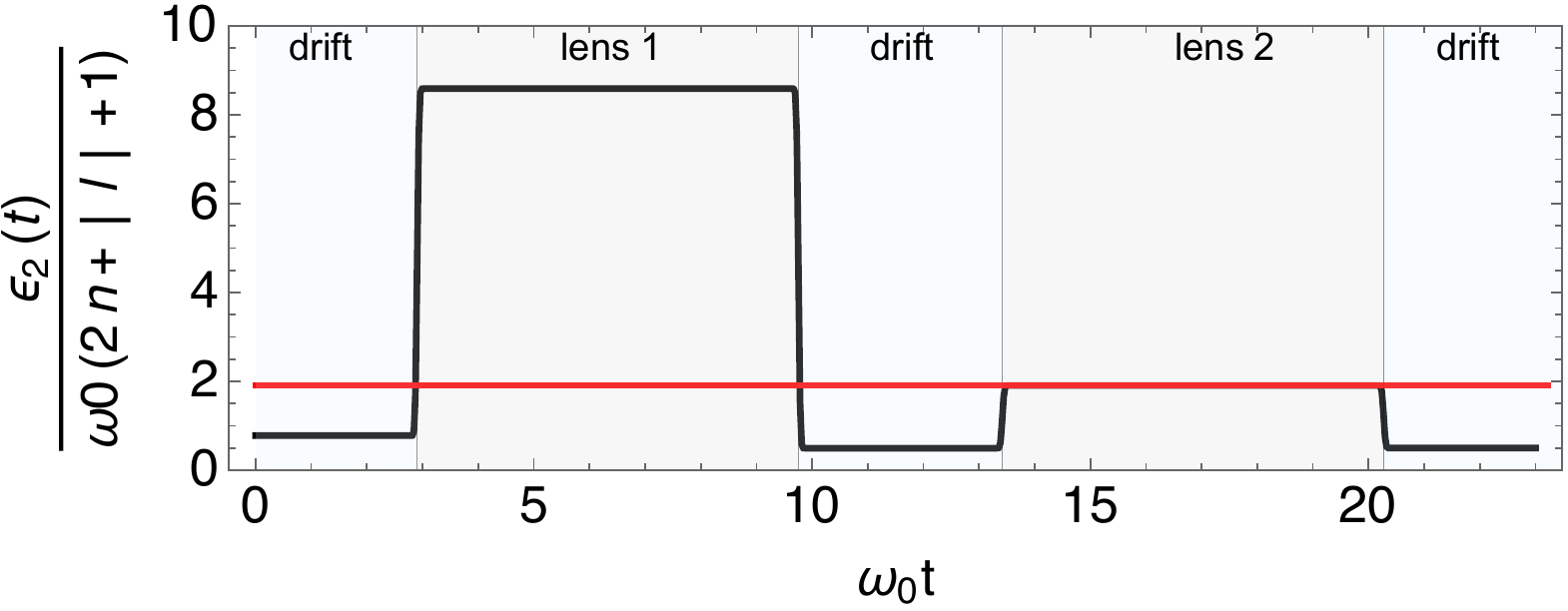}}
\caption{Time evolution of the scaling parameter $b(t)$ (middle panel) and normalized transverse part of the energy of the twisted electron state (bottom panel) in the time varying magnetic field $F(t)$ (upper panel). The red line in the middle panel and in the bottom panel represents scaling parameter $b=1$ and scaled energy Eq.\eqref{eq:LandauE} $\frac{\varepsilon_\perp}{\omega_0(2n+|l|+1)}$ for the stationary Landau state and serves as a reference. For this example we chose $n=0$ and $l=10$.}
\label{Fig:2}
\end{figure}

We consider the propagation of the twisted electron through two consequent solenoids. We assume the electron is twisted, non-relativistic, and is moving along the $z$ axis at a speed $V$. We assume localization of the electron in the longitudinal direction (the longitudinal density is $\rho_z \propto \delta(z-Vt)$). The latter is always possible when the longitudinal momentum dispersion is small $\Delta p_z/\langle p_z \rangle\ll1$ (see Ref.\cite{Baturin} for details on the approximation). The problem has an exact solution with the transverse motion that factors out. The corresponding setup in the transverse plane corresponds to the Eq.\eqref{eq:mainSch}. After simplifications, the final equation that describes transverse motion reads
\begin{align}
\label{eq:NST}
    &i\partial_t \psi_2= \\ \nonumber &\left[\frac{\left[\hat{p}_x^2+\hat{p}_y^2\right]}{2m}+\frac{m\omega^2(t) (\hat{x}^2+\hat{y}^2)}{2}+\omega(t) \hat{L}_z\right]\psi_2, 
\end{align}
with $\omega(t)$ given by the Eq.\eqref{eq:frt}. As before, we consider Landau model Eq.\eqref{eq:DAU} as a reference system. With the help of the Ermakov operator Eq.\eqref{eq:EP2dc} and the expression for the Landau wave-function Eq.\eqref{eq:Landau} one may write the solution to the Eq.\eqref{eq:NST} as
\begin{align}
\label{eq:NSTwf}
    &\psi_2(\rho,\phi,t) = \\ \nonumber &\frac{N}{b} \left(\frac{\rho}{b\rho_H}\right)^{|l|}\mathcal{L}_{n}^{|l|}\left[\frac{2\rho^2}{b^2\rho_H^2}\right]\exp\left[-\frac{\rho^2}{b^2\rho_H^2} \right] \times  \\ \nonumber &\exp\left[-il\int_0^{t}\omega(t')dt' + i l \phi+{\frac{im}{2}\frac{\dot{b}}{b}\rho^2}-i (\varepsilon_\perp-\omega_0 l) \int \frac{dt}{b^2}\right].
\end{align}
Scaling parameter $b$ according to the mapping prescription must satisfy Eq.\eqref{eq:EPeq} with $\gamma_2=0$, $w_1=w_2=1$ and $\omega_1=\omega_0=const$:
\begin{align}
\label{eq:EPeqNSTl}
    \ddot{b} + \omega^2(t) b = \frac{\omega^2_0}{b^3}.
\end{align}
Eq.\eqref{eq:EPeqNSTl} must be complimented with the initial conditions defined by the initial state of the twisted electron. We assume that at $t=0$, transverse mean square radius of the electron  wave packet is $\langle \rho^2 \rangle(0)=0.8^2 \rho_H^2 (2n+|l|+1)$ and $\partial_t\langle \rho^2 \rangle(0)=0$. This corresponds to the focal point in a free space. Initial conditions on the mean square radius translate into the initial conditions for the scaling parameter $b$ as follows
\begin{align}
\label{eq:incond}
&b(0)=0.8, \nonumber \\
&\dot{b}(0)=0.
\end{align} 
To solve Eq.\eqref{eq:EPeqNSTl} numerically, we will need to use the initial conditions Eq.\eqref{eq:incond} and the magnetic field variation shown in Fig.\ref{Fig:2} upper panel. We introduce a dimensionless function $F(t)$ defined as $\omega^2(t)/\omega^2_0$. The magnetic field consists of several segments, starting from free space, a solenoid lens, another free space segment, a second solenoid lens, and a final third free space segment. With these parameters in place, we can proceed with our numerical solution.

As we can see from the middle panel of Fig.\ref{Fig:2}, the wave packet parabolically expands in free space -  a well-known fact in quantum optics. However, inside the first solenoid, the scaling parameter oscillates, leading to oscillations of all wave-packet observables and an increase in state energy, as shown in the bottom panel of Fig.\ref{Fig:2}. This results in time-dependent non-vanishing multipole moments. The lowest quadrupole moment, which is proportional to the average square distance from the origin, is given by 
\begin{align}
Q\propto \langle \rho^2 \rangle \propto b^2(t).
\end{align}   
This in-turn must result in the intensive radiation as 
\begin{align}
\partial_t \varepsilon \propto (\partial^3_t Q(t))^2 \propto (\partial^3_t b^2(t))^2\neq0.  
\end{align}
The latter is the distinct feature of the charged particles only. 
  
 The packet divergence in free space prevents it from being directly captured into the Landau state. The packet size $\sqrt{\langle \rho^2 \rangle(t)}$ oscillates in all practical cases, which means that under the assumptions of the model, a charged quantum wave packet will always radiate when entering a magnetic solenoid.

 Another observation that follows from Fig.\ref{Fig:2} is that time average of $\langle \rho^2 \rangle(t) $ over one period of oscillations $T=2\pi/\omega_0$ is alway greater than the corresponding value $\langle \rho_1^2 \rangle_1$ of the Landau state given by Eq.\eqref{eq:rhost}, as
 \begin{align}
 \frac{1}{2\pi}\int_0^{2\pi} b^2(\tau)d\tau>1    
 \end{align}
 with $\tau=\omega_0 t$. 
 
 However, the geometric average 
 \begin{align}
 \max\left[{b}(t)\right]\min\left[{b}(t)\right]=1,
 \end{align}
 always holds inside the lens. 
 
 This results in
 \begin{align}
\sqrt{ \max\left[\langle \rho^2 \rangle(t)\right]\min\left[\langle \rho^2 \rangle(t)\right]}=\frac{1}{m \omega_0}\left(2n+|l|+1\right),
 \end{align}
 an invariant geometric size of the packet.
 
It is worth mentioning that the selection of appropriate parameters plays a crucial role in capturing free twisted electrons into the Landau state. As evident from Fig.\ref{Fig:2}, placing a second lens at the right spot enables a smooth transition from the vacuum to the Landau level.
  
 It is important to keep in mind that the direct transition to the Landau state is never exact and can only be achieved at one specific point in the parameter space. As a result, $b(t)$ will always oscillate within the lens in any practical setup. However, the amplitude of these oscillations can be minimized, as shown in Fig.\ref{Fig:2}. Additionally, it's important to note that these oscillations will result in radiation as all the multipole moments oscillate while the electron is in the magnetic field. This radiation induced by the oscillations is different from the common spontaneous emission and is most likely classical. However, further investigation and evidence are needed to fully understand this phenomenon.

 One of the immediate consequence of the QAT formalism and the Ermakov mapping is the Ermakov-Lewis invariant \cite{Lewis,Lewis1,Lewis2,NUF21} that for the Hamiltonian Eq.\eqref{eq:NST} after the normalization $\omega_0  \to 1$ and $m \to 1$ is just the sum of the 1D invariants
 \begin{align}
     \hat{I}&=\hat{I}_x+\hat{I}_y=\\&\nonumber \frac{1}{2b^2}\left[\hat{x}^2+(b^2 \hat{p}_x-b\dot{b}\hat{x})^2 \right]+
             \frac{1}{2b^2}\left[\hat{y}^2+(b^2 \hat{p}_y-b\dot{b}\hat{y})^2 \right].
 \end{align}
 We note that classical counterpart of the Ermakov-Lewis invariant is the Courant-Snyder invariant \cite{CS,SYL}: 
 \begin{align}
     \epsilon_x=&\frac{1}{\beta}\left[x^2+(\beta p_x+\alpha x)^2 \right],
 \end{align}
 a quadratic invariant that can aid interpretation of the dynamics in the classical limit.
 
 Courant-Snyder invariant up to a $\pi$ multiplier equals the area of the phase space ellipse of the system and usually referred to as the geometric emittance. 
  Here $\beta$ is the so called envelope function or Twiss beta-function; $\alpha$ - is the Twiss alpha-function that is connected to $\beta$ as
 \begin{align}
     \alpha=-\frac{\dot{\beta}}{2}.
 \end{align}
 
 Geometric emittance also equals twice the action. We note that the Ermakov-Lewis invariant is the action. To bridge these two quantities we introduce a quantum emittance operator as:
\begin{align}
    &\hat{\epsilon}_{x,y}=2\hat{I}_{x,y}, \\&\hat{\epsilon}_{x}=\frac{1}{b^2}\left[\hat{x}^2+(b^2 \hat{p}_x-b\dot{b}\hat{x})^2 \right], \nonumber \\&\hat{\epsilon}_{y}=\frac{1}{b^2}\left[\hat{y}^2+(b^2 \hat{p}_y-b\dot{b}\hat{y})^2 \right].  \nonumber
\end{align}
Dimensionless Twiss beta-function $\beta$ is connected to $b$ as
\begin{align}
    \beta=b^2
\end{align}
and $\alpha$ parameter is just
\begin{align}
    \alpha=-b\dot{b}.
\end{align}
We point out that the quantum emittance as well as it's variance is conserved and defined by the initial conditions and the quantum state only.  
 
 
 
 
 
 \subsection{Non-stationary current operator}
 
It is convenient to recast the current operator in terms of the current operator for the stationary Landau state, to analyze the coupling of the non-stationary states to the quantized external electromagnetic field. This is extremely helpful when it comes to the analysis of the possible decay channels of the general non-stationary states to the stationary Landau states as well as to study different scattering processes that involves twisted particles (see Ref.\cite{IvanovPubl} for the details). 

One may use known formulas for the corresponding integrals for the stationary Landau states (see, for instance, Ref.\cite{ST} and some recent results in Ref.\cite{EPP}) when calculating the S-matrix with non-stationary states.  

First, we consider a general case of a non-stationary magnetic field with a dissipation guided by Eq.\eqref{eq:CKH}. The transverse part of the vector potential for this model has the form
\begin{align}
    \vec{A}_2^{T}=\left\{-\frac{B(t_2) y_2}{2 \sqrt{w_2}}, \; \frac{B(t_2) x_2}{2 \sqrt{w_2}} \right\}.
\end{align}
As an initial system for the mapping, we chose the Landau model guided by Eq.\eqref{eq:DAU}. 
According to the transformation, Eq.\eqref{eq:EP2dc}, and expression for the transformed momentum, Eq.\eqref{eq:mom_tr}, the potential for the gauge field is
\begin{align}
\label{eq:Gpot}
    \vec{G}=\frac{m}{w_2}\frac{\dot{b}}{b}\vec{r}_1.
\end{align}
The probability current is given by
\begin{align}
\label{eq:cur}
    \vec{\hat{j}_2} =  \frac{\Re{\psi^{*}_2 \sqrt{w_2}\vec{\hat{p}_2} \psi_2}}{m} - \frac{e}{m}\vec{A}_2|\psi_2|^2 .
\end{align}
Taking into account that 
\begin{align}
\label{eq:cur2}
    - e \vec{A_2} |\psi_2|^2 = - e \frac{1}{\sqrt{w_2}} \frac{B(t_2)}{b B_0} \vec{A_1} |\psi_1|^2,
\end{align}
where $\vec{A}_1^T=\left(-\frac{B_0 y_1}{2}, \; \frac{B_0 x_1}{2}\right)$ 
is the transverse part of the vector potential for the Landau system, we combine  
Eq.\eqref{eq:Gpot}, Eq.\eqref{eq:mom_tr} and Eq.\eqref{eq:cur} to achieve
\begin{align}
\label{eq:trcur}
    &\vec{\hat{j}_2} = \\ \nonumber&\frac{\sqrt{w_2}}{b^2}\left\{\frac{\vec{\hat{j}_1}}{b}  +\left[\left(\frac{1}{b}-\frac{B(t_2)b}{B _0w_2} \right) \frac{e\vec{A_1}}{m}+\frac{b}{m} \vec{G}\right]|\psi_1|^2\right\}.
\end{align}
For the case of the non-stationary Landau states one should set $w_2=1$, $B(t_2)=B_0$, and $b$ must satisfy Eq.\eqref{eq:EPeqDau}. We note that probability current is not invariant under the Ermakov transformation. However, the structure of $\vec{\hat{j}_2}$ is simple. Namely, it consists of the re-scaled original current, $\vec{\hat{j}_1}$, and two vector potentials that are linear functions of the coordinates. We highlight that spatial integrals over $d x_2 dy_2$ could be evaluated using stationary states only. This opens a simple and powerful method for the calculation of scattering amplitudes.   

\section{Conclusion}

We have discussed a QAT approach to the time solution of Schr\"{o}dinger's equation within the Cardirola-Kanai model. This approach takes into account the classical friction in a quantum system. The QAT formalism has been applied to analyze the evolution of the general twisted quantum state based on the corresponding stationary state. We have explicitly derived the 2D Ermakov mapping given by Eq.\eqref{eq:EP2d} and Eq.\eqref{eq:EP2dc}, which provides a unique correspondence between the stationary twisted state and its evolving counterpart. We reiterate that the unitary property of the 2D Ermakov operator leads to the fact that all dynamic twisted states, including free states, non-stationary states, and states that may appear in different time-dependent fields, are unitarily equivalent to the corresponding stationary states. We illustrated the method by a trivial derivation of the non-stationary Landau state and provided an explicit formula for the energy of this state. We established a connection between a classical motion of the electron and a quantum electron wave-packet in the time-dependent magnetic field and introduced the wave-packet emittance operator. Finally, we derived the current operator for the non-stationary system, which provides a convenient way to analyze the coupling to the quantized external fields of the non-stationary twisted state.

In conclusion, we emphasize that the 2D Ermakov mapping given by Eq.\eqref{eq:EP2d} and Eq.\eqref{eq:EP2dc} is valid for any wave function (not necessarily eigenstates of the Hamiltonian) and applies to the analysis of the dynamics of the wide class of wave packet solutions, including non-pure states. Once combined with the Foldy-Wouthuysen transformation \cite{FW,SilenkoFW}, it can become a convenient tool for a relativistic packet state description. The Ermakov mapping naturally accounts for radiation friction, which becomes essential in problems related to the generation of high-energy twisted electrons in linear accelerators.

\begin{acknowledgments}
The work is funded by Russian Science Foundation and St. Peterburg Science Foundation, project № 22-22-20062, https://rscf.ru/project/22-22-20062/. The authors are grateful to Ivan Terekhov, Igor Chestnov and Gerard Andonian for useful discussions and suggestions.
\end{acknowledgments}

\begin{widetext}

\appendix
 
\section{The Arnold Transformation.}

\label{QATcl}

We consider two classical systems, a linear oscillator with linear friction and a free system \cite{QAT1}. The Euler-Lagrange equation of the oscillator is 
\begin{align}
\label{eq:osc}
    \ddot{x} + \dot{f} \dot{x} + \omega^2 x = 0,
\end{align}
and for the free particle
\begin{align}
    \stackrel{\circ \circ}{\kappa} \; = 0.
\end{align}
The dot above denotes the total derivative by the time $t$ 
\begin{align}
    \frac{df}{dt}\equiv \dot{f},
\end{align}
and circle above denotes the total derivative by the time $\tau$
\begin{align}
    \frac{dg}{d\tau}\equiv ~\stackrel{\circ}{g}.
\end{align}
A local map that establishes a connection between these two systems is given by the classical Arnold transformation discussed in \cite{QAT1}:
\begin{align*}
    \mathds{R} \times T \rightarrow \mathds{R} \times \mathcal{T}, \\
    (x, t) \mapsto (\kappa, \tau),
\end{align*}
and 
\begin{equation}
\label{eq:CAT}
\begin{cases}
    \kappa = \frac{x}{u_2}, \\
   \tau = - \frac{u_1}{u_2}.
\end{cases}
\end{equation}
Where $u_1, u_2$ are two linearly independent solutions of the equation.\eqref{eq:osc}.\\
We notice the following connection between the time derivatives
 \begin{align}
 \label{eq:def}
     \dot{\tau} = - \frac{\dot{u}_1 u_2 - \dot{u}_2 u_1}{u^2_2} = \frac{w}{u^2_2} \; &\Rightarrow \;\; \stackrel{\circ}{t} \; = \frac{u^2_2}{w}. 
 \end{align}
Where $w \equiv u_1\dot u_2 - \dot u_1 u_2 $ - is the Wronskian built on $u_1$ and $u_2$. For the Wronskian $w$ the following is true:
 \begin{align}
 \label{eq:dwronsk}
    w = e^{-f} \; &\Rightarrow \; \dot{w} = - \dot{f} w.
 \end{align}
We make a local change of variables explicitly as follows
\begin{align}
    0 &= \ddot{x} + \dot{f} \dot{x} + \omega^2 x = \ddot{\kappa} u_2 + 2 \dot{\kappa} \dot{u}_2 + \kappa \ddot{u}_2 + \dot{f} \dot{\kappa} u_2 + \dot{f} \kappa \dot{u}_2 + \omega^2 \kappa u_2 = \nonumber\\
    &= \frac{d}{dt}\left(\stackrel{\circ}{\kappa} \frac{w}{u^2_2}\right) u_2 + 2 \stackrel{\circ}{\kappa} \frac{w}{u^2_2} \dot{u}_2 + \dot{f} \stackrel{\circ}{\kappa} \frac{w}{u^2_2} u_2 + \kappa (\uwave{\ddot{u}_2 + \dot{f} \dot{u}_2 + \omega^2 u_2}) = \nonumber\\
    &= \stackrel{\circ \circ}{\kappa} \frac{w^2}{u^4_2} u_2 \; \uline{- \; 2 \stackrel{\circ}{\kappa} \frac{w}{u^3_2} \dot{u}_2 u_2} + \stackrel{\circ}{\kappa} \frac{\dot{w}}{u_2} \; \uline{+ \;2 \stackrel{\circ}{\kappa} \frac{w}{u^2_2} \dot{u}_2} + \dot{f} \stackrel{\circ}{\kappa} \frac{w}{u_2} = \\
    &= \stackrel{\circ \circ}{\kappa} \frac{w^2}{u^3_2} \; \uuline{-\; \stackrel{\circ}{\kappa} \frac{\dot{f} w}{u_2}} \; \uuline{+ \; \dot{f} \stackrel{\circ}{\kappa} \frac{w}{u_2}} = \stackrel{\circ \circ}{\kappa} \frac{w^2}{u^3_2}. \nonumber
\end{align}
Thus we observe that one system is mapped to another under the coordinate substitution Eq.\eqref{eq:CAT} up to a multiplier $w^2/u^3_2$. The idea behind the transformation is quite simple and goes back (according to Arnold) to Newton \cite{Arnold}. The locality of the transformation follows from the coordinate substitution, which is inversely proportional to $u_2$. Consequently, the applicability condition is $u_2 \neq 0$. This implies that if $u_2$ is periodic, then the transformation holds only within one period.  

\section{The Quantum Arnold Transformation}
\label{QATq}

Along with the classical transformation, one may introduce its Quantum analog that maps the Hilbert space of solutions of the Schr\"{o}dinger's equation for the quantum oscillator $\mathcal{H}_t$  to the Hilbert space of solutions $\mathcal{H}_\tau$ of the Schr\"{o}dinger's equation for the free particle \cite{QAT1}. 

It is apparent that coordinate substitution Eq.\eqref{eq:CAT} generates the change in the momentum given by
\begin{align}
    p_\tau = m \frac{d \kappa}{d\tau}= \frac{\stackrel{\circ}{t}}{u_2} p_x-m \frac{\dot{u_2}}{u_2^2} \stackrel{\circ}{t} x,
\end{align}
consequently, the gauge potential \cite{davidov} is 
\begin{align}
    G=-m \frac{\dot{u_2}}{u_2^2} \stackrel{\circ}{t} x=-m \frac{\dot{u_2}}{w}x.
\end{align}
On the other hand 
\begin{align}
    G=\frac{d \theta_\kappa}{d \kappa},
\end{align}
where $\theta_\kappa$ is the phase of the wave function. \\
Therefore, the phase of the wave-function induced by the transformation Eq.\eqref{eq:CAT}  is 
\begin{align}
    \theta_\kappa=-m \frac{\dot{u_2}}{w} \int\limits_{0}^x \tilde{x} d\kappa=-\frac{m}{w} \frac{\dot{u_2}}{u_2} \int\limits_{0}^x \tilde{x} d \tilde{x}=-\frac{m}{2w} \frac{\dot{u_2}}{u_2} x^2.
\end{align}
Accounting for the change in norm induced by the coordinate substitution one may introduce the Quantum Arnold Transformation (QAT) as  
\begin{equation}
\label{eq:QATa}
\hat{\mathcal{Q}}:
\begin{cases}
    \kappa = \frac{x}{u_2}, \\
    \tau = - \frac{u_1}{u_2}, \\
    \varphi(\kappa, \tau) = \psi(x, t) \sqrt{u_2} e^{- \frac{i}{2} \frac{m}{w} \frac{\dot{u}_2}{u_2}x^2}.
\end{cases}
\end{equation}
QAT provides the mapping of one Schr\"{o}dinger's equation to another as:
\begin{align*}
    i \frac{\partial \varphi}{\partial \tau} = - \frac{1}{2 m} \frac{\partial^2 \varphi}{\partial \kappa^2} \; \Rightarrow \; i \frac{\partial \psi}{\partial t} = - \frac{w}{2 m} \frac{\partial^2 \psi}{\partial x^2} + \frac{m \omega^2 x^2}{2 w} \psi.
\end{align*}
Explicitly the wave-function of the free system $\varphi(\kappa, \tau)$ reads: 
\begin{align*}
     \varphi(\kappa, \tau) = \psi\big(\kappa u_2(t(\tau)), t(\tau)\big) \sqrt{u_2(t(\tau))} \; e^{- \frac{i}{2} \frac{m}{w(t(\tau))} \dot{u}_2(t(\tau)) u_2(t(\tau)) \kappa^2}.
\end{align*}
Next, we write left and right hand side of the Schr\"{o}dinger's equation in the expended form and proceed with the substitution \eqref{eq:QATa}.
The left-hand side reads
\begin{align*}
     &i \frac{\partial \varphi}{\partial \tau} = \Bigg[i \frac{\partial \psi}{\partial x} \kappa \dot{u}_2 \stackrel{\circ}{t} \sqrt{u_2} +  i \frac{\partial \psi}{\partial t} \stackrel{\circ}{t} \sqrt{u_2} +  i \psi \frac{\dot{u}_2 \stackrel{\circ}{t}}{2 \sqrt{u_2}} + \\
     &+ i \psi \sqrt{u}_2 \left\{- \frac{i}{2} m \kappa^2 \stackrel{\circ}{t} \left(\frac{\ddot{u}_2 u_2 + \dot{u}^2_2}{w} - \frac{\dot{w}}{w^2} \dot{u}_2 u_2 \right) \right\} \Bigg] e^{- \frac{i}{2} \frac{m}{w} \dot{u}_2 u_2 \kappa^2} = \Bigg[\uline{i \frac{\partial \psi}{\partial x} \kappa \frac{\dot{u}_2 u^2_2}{w} \sqrt{u_2}}\; + \\
     &+ i \frac{\partial \psi}{\partial t} \frac{u^2_2}{w} \sqrt{u_2} \; \uuline{+ \; i \psi \frac{\dot{u}_2 u^2_2}{2 w \sqrt{u_2}}} + i \psi \sqrt{u}_2 \left\{-\frac{i}{2} m \kappa^2 \frac{u^2_2}{w^2} \left(- \omega^2 u^2_2 + \uwave{\dot{u}^2_2} \right) \right\} \Bigg] e^{-\frac{i}{2} \frac{m}{w} \dot{u}_2 u_2 \kappa^2}.
\end{align*}
The right-hand side reads
\begin{align*}
    - \frac{1}{2 m} \frac{\partial^2 \varphi}{\partial \kappa^2} &= \Bigg[ - \frac{1}{2 m} \frac{\partial^2 \psi}{\partial x^2} u^2_2 \sqrt{u_2} \; \uline{- \; \frac{1}{m} \frac{\partial \psi}{\partial x} u_2 \sqrt{u_2} \left\{- i  \frac{m}{w} \dot{u}_2 u_2 \kappa \right\}} - \\
    &- \frac{1}{2 m} \psi \sqrt{u_2} \left\{\uuline{- i m \dot{u}_2 u_2 \frac{1}{w}} \; \uwave{- \;\frac{m^2}{w^2} \dot{u}^2_2 u^2_2 \kappa^2} \right\} \Bigg] e^{-\frac{i}{2} \frac{m}{w} \dot{u}_2 u_2 \kappa^2}.
\end{align*}
After cancellation of the common terms on both side we arrive at 
\begin{align*}
    i \frac{\partial \psi}{\partial t} \frac{u^2_2}{w} \sqrt{u_2} - \frac{m \omega^2 x^2}{2 w^2} \psi u^2_2 \sqrt{u}_2 = - \frac{1}{2 m} \frac{\partial^2 \psi}{\partial x^2} u^2_2 \sqrt{u}_2.
\end{align*}
Thus the free Schr\"{o}dinger's equation under the inverse QAT transforms to the Schr\"{o}dinger's equation of the quantum harmonic oscillator up to a common multiplier $\frac{u^2_2 \sqrt{u_2}}{w}$.

\section{A one dimensional Ermakov mapping}
\label{Ermakov1D}

We consider the following mapping diagram \cite{QAT2,QAT3}:
\begin{align*}
    \mathcal{H}_1 \xrightarrow{\;\;\;\;\;\hat{\mathcal{Q}}_2^{-1}\hat{\mathcal{Q}}_1\;\;\;\;\;} \mathcal{H}_2 \\
    \hat{\mathcal{Q}}_1 \searrow \;\;\;\;\;\;\; \swarrow \hat{\mathcal{Q}}_2 \;\; \\
    \mathcal{H}_\text{free} \;\;\;\;\;\;\;\;\;\;
\end{align*}
where $\mathcal{H}_\text{free}, \mathcal{H}_1$ and $\mathcal{H}_2$ - are Hilbert spaces that correspond to the free quantum particle and two 1D quantum harmonic oscillators with frequencies $\omega_1$ and $\omega_2$ correspondingly.
It is apparent that the mapping 
\begin{align*}
    \hat{\mathcal{Q}}^{-1}_2 \hat{\mathcal{Q}}_1: \;\;\;\;\;\;\;\;\;\; &\mathcal{H}_1 \longrightarrow \mathcal{H}_2, \\
    \psi_1(x_1&, t_1) \mapsto \psi_2(x_2, t_2),
\end{align*}
maps one oscillator on to another. 
Classical Euler-Lagrange equation for the system 1 that has a Hilbert space of solutions $\mathcal{H}_1$ is
\begin{align}
\label{eq:osc1}
    \stackrel{\circ \circ}{x}_1 + \stackrel{\circ}{f}_1 \stackrel{\circ}{x}_1 + \;\omega^2_1 x_1 = 0.
\end{align} 
For the two linearly independent solutions of this equation we adopt the following notation $u^{(1)}_1$, $u^{(1)}_2$ and Wronskian for these solutions is $w_1 = u^{(1)}_1 \dot{u}^{(1)}_2 - u^{(1)}_2 \dot{u}^{(1)}_1 = e^{-f_1}$. One may check that the following is true (by analogy to  \eqref{eq:dwronsk}): $\stackrel{\circ}{w}_1 = - \stackrel{\circ}{f}_1 w_1$. \\
Classical Euler-Lagrange equation for the system 2 with the Hilbert space of solutions $\mathcal{H}_2$ is
\begin{align}
\label{eq:osc2}
    \ddot{x}_2 + \dot{f}_2 \dot{x}_2 + \omega^2_2 x_2 = 0.
\end{align} 
We denote by $u^{(2)}_1$ and $u^{(2)}_2$ two linear independent solutions of the Eq.\eqref{eq:osc2}, and  $w_2 = u^{(2)}_1 \dot{u}^{(2)}_2 - u^{(2)}_2 \dot{u}^{(2)}_1 = e^{-f_2}$ - the Wronskian. Similarly to the previous case $\dot{w}_2 = - \dot{f}_2 w_2$. \\
We introduce new function $b(t_2) = \frac{u^{(2)}_2}{u^{(1)}_2} = \frac{x_2 k}{x_1 k} = \frac{x_2}{x_1}$.\\
From Eq.\eqref{eq:QATa} for $\hat{\mathcal{Q}}_1$ and $\hat{\mathcal{Q}}_2$ one may derive: 
\begin{align*}
    \frac{w_1(t_1)}{\left(u^{(1)}_2\right)^2} dt_1 = d \tau = \frac{w_2(t_2)}{\left(u^{(2)}_2\right)^2} dt_2 \; \Rightarrow \; \dot{t}_1 = \frac{w_2(t_2)}{w_1(t_1)} \frac{1}{b^2(t_2)}.
\end{align*}
Next we derive differential equation for $b(t_2)$:
\begin{align}
    0 &= \ddot{x}_2 + \dot{f}_2 \dot{x}_2 + \omega^2_2 x_2 = x_1 \ddot{b} + 2 \dot{x}_1 \dot{b} + \ddot{x}_1 b + \dot{f}_2 x_1 \dot{b} + \dot{f}_2 \dot{x}_1 b + \omega^2_2 x_1 b = x_1 \ddot{b} \; \uline{+ \; 2 \stackrel{\circ}{x}_1 \frac{w_2}{w_1} \frac{1}{b^2} \dot{b}} \; + \nonumber \\
    &+ \stackrel{\circ \circ}{x}_1 \frac{w^2_2}{w^2_1} \frac{b}{b^4} + \stackrel{\circ}{x}_1 \frac{\dot{w}_2 b}{w_1 b^2} \; \uline{- \; 2 \stackrel{\circ}{x}_1 \frac{w_2 \dot{b} b}{w_1 b^3}}\; - \stackrel{\circ}{x}_1 \frac{w_2 \dot{w}_1 b}{w^2_1 b^2} + \dot{f}_2 x_1 \dot{b} + \dot{f}_2 \stackrel{\circ}{x}_1 \frac{w_2}{w_1} \frac{1}{b} + \omega^2_2 x_1 b = \nonumber \\
    &= x_1 \ddot{b} \; \uuline{- \stackrel{\circ}{f}_1 \stackrel{\circ}{x}_1 \frac{w^2_2}{w^2_1} \frac{1}{b^3}} - \omega^2_1 x_1 \frac{w^2_2}{w^2_1} \frac{1}{b^3} \; \uwave{- \stackrel{\circ}{x}_1 \dot{f}_2 \frac{w_2}{w_1 b}} \; \uuline{+ \stackrel{\circ}{x}_1 \frac{w^2_2}{w^2_1 b^3} \stackrel{\circ}{f}_1} + \dot{f}_2 x_1 \dot{b} \; \uwave{+ \; \dot{f}_2 \stackrel{\circ}{x}_1 \frac{w_2}{w_1} \frac{1}{b}} + \omega^2_2 x_1 b = \nonumber \\
    &= x_1 \ddot{b} - \frac{w^2_2}{w^2_1} \frac{\omega^2_1}{b^3} x_1 + \dot{f}_2 x_1 \dot{b} + \omega^2_2 x_1 b \; \Rightarrow \; \boxed{\ddot{b} + \dot{f}_2 \dot{b} + \omega^2_2 b = \frac{w^2_2}{w^2_1} \frac{\omega^2_1}{b^3}}.
\end{align}
The last equation is a well-known Ermakov-Pinney equation \cite{Ermakov,Pinney}. The mapping of one oscillator onto another is possible if and only if the $b$ mapping parameter satisfies the Ermakov-Pinney equation as given above. \\
Combining direct and inverse QAT for the two systems one can build a mapping that maps one system onto another in a sense of Hilbert spaces. Explicitly this transformation is given by the Ermakov operator and reads
\begin{equation}
\label{eq:QAT2}
\hat{\mathcal{E}}_{1 \mapsto 2} = \hat{\mathcal{Q}}_2^{-1}\hat{\mathcal{Q}}_1:
\begin{cases}
    x_2 = b x_1, \\
   dt_2 = \frac{w_1(t_1)}{w_2(t_2)} b^2(t_2) dt_1,\\
    \psi_2(x_2,t_2) = \psi_1(x_1,t_1) \frac{1}{\sqrt{b}} e^{\frac{i}{2} \frac{m}{w_2} \frac{\dot{b}}{b} x^2_2}.
\end{cases}
\end{equation}
It consists of coordinate and time substitution and a phase multiplier due to the gauge transformation. \\
This mapping enables the following transition:
\begin{align*}
    i \frac{\partial \psi_2}{\partial t_2} = -\frac{w_2}{2 m} \frac{\partial^2 \psi_2}{\partial x^2_2} + \frac{m \omega^2_2 x^2_2}{2 w_2} \psi_2 \; \Rightarrow \; i \frac{\partial \psi_1}{\partial t_1} = -\frac{w_1}{2 m} \frac{\partial^2 \psi_1}{\partial x^2_1} + \frac{m \omega^2_1 x^2_1}{2 w_1} \psi_1,
\end{align*}
Next, we show this transition explicitly.
The left-hand side reads
\begin{align*}
    i \frac{\partial \psi_2}{\partial t_2} &= \left[ - i \frac{\partial \psi_1}{\partial x_1} \frac{x_2 \dot{b}}{b^2} \frac{1}{\sqrt{b}} + i \frac{\partial \psi_1}{\partial t_1} \dot{t}_1 \frac{1}{\sqrt{b}} - i \psi_1 \frac{\dot{b}}{2 b \sqrt{b}} - \frac{m}{2} \psi_1 \frac{x^2_2}{\sqrt{b}}\left\{-\frac{\dot{w}_2}{w^2_2} \frac{\dot{b}}{b} + \frac{1}{w_2} \frac{\ddot{b} b - \dot{b}^2}{b^2} \right\} 
    \right] e^{\frac{i}{2} \frac{m}{w_2} \frac{\dot{b}}{b} x^2_2} = \\
    &= \Bigg[\uline{- i \frac{\partial \psi_1}{\partial x_1} \frac{x_2 \dot{b}}{b^2} \frac{1}{\sqrt{b}}}
    + i \frac{\partial \psi_1}{\partial t_1} \frac{w_2}{w_1 b^2} \frac{1}{\sqrt{b}} \; \uline{\uline{-\; i \psi_1 \frac{\dot{b}}{2 b \sqrt{b}}}} - \frac{m}{2} \psi_1 \frac{x^2_2}{w_2 b \sqrt{b}} \times \\
    &\times \left\{\uwave{\frac{\dot{f}_2 w_2 \dot{b}}{w_2}} \; \uwave{- \dot{f}_2 \dot{b}} \; \uwave{\uwave{-\;\omega^2_2 b}} + \frac{w^2_2}{w^2_1} \frac{\omega^2_1}{b^3} \;\uuline{-\; \frac{\dot{b}^2}{b}} \right\} 
    \Bigg] e^{\frac{i}{2}  \frac{m}{w_2} \frac{\dot{b}}{b} x^2_2}.
\end{align*}
The right-hand side reads
\begin{align*}
    -\frac{w_2}{2 m} \frac{\partial^2 \psi_2}{\partial x^2_2} &= \Bigg[-\frac{w_2}{2 m} \frac{\partial^2 \psi_1}{\partial x^2_1} \frac{1}{b^2 \sqrt{b}} \; \uline{- \; \frac{i}{2} \frac{\partial \psi_1}{\partial x_1} \frac{1}{b \sqrt{b}} \frac{\dot{b}}{b} 2 x_2} - \frac{i}{2} \psi_1 \frac{\dot{b}}{b \sqrt{b}} \left\{\uline{\uline{1}} \;\uuline{+ \; \frac{i}{2} \frac{m}{w_2} \frac{\dot{b}}{b} 2 x^2_2} \right\}\Bigg] e^{\frac{i}{2} \frac{m}{w_2} \frac{\dot{b}}{b} x^2_2}\\ \\
     \frac{m \omega^2_2 x^2_2}{2 w_2} \psi_2 &=  \uwave{\uwave{\frac{m \omega^2_2 x^2_2}{2 w_2} \psi_1 \frac{1}{\sqrt{b}} e^{\frac{i}{2} \frac{m}{w_2} \frac{\dot{b}}{b} x^2_2}}}
\end{align*}
After cancellation of the common terms on both side we arrive at 
\begin{align*}
    i \frac{\partial \psi_1}{\partial t_1} \frac{w_2}{w_1 b^2} \frac{1}{\sqrt{b}} - \frac{m}{2} \frac{w_2}{w^2_1} \frac{\omega^2_1 x^2_1}{b^2 \sqrt{b}} \psi_1 = -\frac{w_2}{2 m} \frac{\partial^2 \psi_1}{\partial x^2_1} \frac{1}{b^2 \sqrt{b}}
\end{align*}
Equation above is exactly the Schr\"{o}dinger's equation for the first system up to a common multiplier  $\frac{w_2}{w_1 b^2 \sqrt{b}}$.

\section{A two dimensional Ermakov mapping}
\label{Erakov2D}

We consider a two dimensional Hamiltonian of the form
\begin{align}
    \hat{H}_2 = \frac{\hat{p}^2_x + \hat{p}^2_y}{2m} w_2 + \frac{m \omega^2_2 (\hat{x}^2 + \hat{y}^2)}{2 w_2} + \omega_2 \hat{L}_z,
\end{align}
along with the 2D Ermakov mapping that reads
\begin{equation}
\label{eq:QAT2D}
\hat{\mathcal{E}}^{2D}_{1 \mapsto 2}:
\begin{cases}
    x_2 = b x_1, \\
    y_2 = b y_1, \\
   dt_2 = \frac{w_1(t_1)}{w_2(t_2)} b^2(t_2) dt_1,\\
    \psi_2(x_2, y_2,t_2) = \frac{1}{b} \psi_1(x_1, y_1, t_1)  e^{\frac{i}{2} \frac{m}{w_2} \frac{\dot{b}}{b} (x^2_2 + y^2_2)}e^{-il_2 \int \omega_2 dt_2 + il_1 \int \omega_1 dt_1}.
\end{cases}
\end{equation}
First we prove that under the mapping \eqref{eq:QAT2D} eigenvalue of the $\hat{L}_z$ operator is conserved $l_1=l_2$. \\
We consider an action of the operator $\hat{L}_{z_2}$ on the wave-function
\begin{align*}
    \hat{L}_{z_2} \psi_2 = (\hat{x}_2 \hat{p}_{y_2} - \hat{y}_2 \hat{p}_{x_2}) \psi_2
\end{align*}
with
\begin{align*}
    \hat{p}_{x_2} \psi_2 = - i \frac{\partial \psi_2}{\partial x_2} = \left[- i  \frac{\partial \psi_1}{\partial x_1} \frac{1}{b^2}  + \frac{m}{w_2} \frac{\dot{b}}{b^2} x_2 \psi_1 \right] e^{\frac{i}{2} \frac{m}{w_2} \frac{\dot{b}}{b} (x^2_2 + y^2_2)}e^{-i l_2 \int \omega_2 dt_2 + i l_1 \int \omega_1 dt_1}
 \end{align*}
 and
  \begin{align*}
     \hat{p}_{y_2} \psi_2 = - i \frac{\partial \psi_2}{\partial y_2} = \left[- i  \frac{\partial \psi_1}{\partial y_1} \frac{1}{b^2}  +  \frac{m}{w_2} \frac{\dot{b}}{b^2} y_2 \psi_1 \right] e^{\frac{i}{2}  \frac{m}{w_2} \frac{\dot{b}}{b} (x^2_2 + y^2_2)}e^{-i l_2 \int \omega_2 dt_2 + i l_1 \int \omega_1 dt_1}
 \end{align*}
 we have
 \begin{align*}
     \hat{L}_{z_2} \psi_2 &= \left[- i x_1 \frac{\partial \psi_1}{\partial y_1} \frac{1}{b} \; \uwave{+ \; \frac{m}{w_2} \dot{b} x_1 y_1 \psi_1} + i y_1 \frac{\partial \psi_1}{\partial x_1} \frac{1}{b} \; \uwave{- \; \frac{m}{w_2} \dot{b} x_1 y_1 \psi_1} \right] e^{\frac{i}{2} \frac{m}{w_2} \frac{\dot{b}}{b} (x^2_2 + y^2_2)} e^{-i l_2 \int \omega_2 dt_2 + i l_1 \int \omega_1 dt_1} = \\
     &= \frac{1}{b} \left[ \hat{L}_{z_1} \psi_1 \right] e^{\frac{i}{2} \frac{m}{w_2} \frac{\dot{b}}{b} (x^2_2 + y^2_2)} e^{-i l_2 \int \omega_2 dt_2 + i l_1 \int \omega_1 dt_1}.
 \end{align*}
 Accounting for the fact that Jacobian determinant of the coordinate transformation is just $b^2$ we have   
 \begin{align*}
     l_2 &= \langle \psi_2| \hat{L}_{z_2} | \psi_2 \rangle = \int{\psi^{*}_2 \hat{L}_{z_2} \psi_2 dx_2 dy_2} = \int{\frac{1}{b} \psi^{*}_1 \frac{1}{b} \hat{L}_{z_1} \psi_1 b^2 dx_1 dy_1} = \langle \psi_1| \hat{L}_{z_1} | \psi_1 \rangle = l_1.
 \end{align*}
This simplifies 2D Ermakov mapping \eqref{eq:QAT2D} to
 \begin{equation}
\label{eq:QAT2D_2}
\hat{\mathcal{E}}^{2D}_{1 \mapsto 2}:
\begin{cases}
    x_2 = b x_1, \\
    y_2 = b y_1, \\
   dt_2 = \frac{w_1(t_1)}{w_2(t_2)} b^2(t_2) dt_1,\\
    \psi_2(x_2, y_2,t_2) = \frac{1}{b} \psi_1(x_1, y_1, t_1)  e^{\frac{i}{2} \frac{m}{w_2} \frac{\dot{b}}{b} (x^2_2 + y^2_2)}e^{-i l \int \omega_2 dt_2 + i l \int \omega_1 dt_1}.
\end{cases}
\end{equation}
2D Ermakov mapping \eqref{eq:QAT2D_2} enables to the following transition 
 \begin{align*}
    i \frac{\partial \psi_2}{\partial t_2} = -\frac{w_2}{2 m} \left(\frac{\partial^2 \psi_2}{\partial x^2_2} + \frac{\partial^2 \psi_2}{\partial y^2_2}\right) + \frac{m \omega^2_2 (x^2_2 + y^2_2)}{2 w_2} \psi_2 + \omega_2 \hat{L}_{z_2} \psi_2 \; \Rightarrow \\
    \Rightarrow \; i \frac{\partial \psi_1}{\partial t_1} = - \frac{w_1}{2 m} \left(\frac{\partial^2 \psi_1}{\partial x^2_1} + \frac{\partial^2 \psi_1}{\partial y^2_1}\right) + \frac{m \omega^2_1 (x^2_1 + y^2_1)}{2 w_1} \psi_1 + \omega_1 \hat{L}_{z_1} \psi_1.
\end{align*}
We show this explicitly below. \\
The left-hand side reads
\begin{align*}
    i \frac{\partial \psi_2}{\partial t_2} &= \Bigg[- i \frac{\partial \psi_1}{\partial x_1} \frac{\dot{b}}{b^3} x_2 - i \frac{\partial \psi_1}{\partial y_1} \frac{\dot{b}}{b^3} y_2 + i \frac{\partial \psi_1}{\partial t_1} \dot{t}_1 \frac{1}{b} - i \psi_1 \frac{\dot{b}}{b^2} - \frac{m}{2} \frac{\psi_1}{b} \left\{\frac{1}{w_2} \frac{\ddot{b}b - \dot{b}^2}{b^2} - \frac{\dot{b}}{b} \frac{\dot{w}_2}{w^2_2}\right\} (x^2_2 + y^2_2) \;+\\
    &+ l \omega_2 \frac{\psi_1}{b} - l \omega_1 \frac{\psi_1}{b} \frac{w_2}{w_1} \frac{1}{b^2} \Bigg] e^{\frac{i}{2} \frac{m}{w_2} \frac{\dot{b}}{b} (x^2_2 + y^2_2)} e^{-il \int \omega_2 dt_2 + il \int \omega_1 dt_1} = \\
    &= \Bigg[\uline{- \; i \frac{\partial \psi_1}{\partial x_1} \frac{\dot{b}}{b^3} x_2 - i \frac{\partial \psi_1}{\partial y_1} \frac{\dot{b}}{b^3} y_2} + i \frac{\partial \psi_1}{\partial t_1} \frac{w_2}{w_1} \frac{1}{b^3} \; \uline{\uline{-\; i \psi_1 \frac{\dot{b}}{b^2}}} - \frac{m}{2} \psi_1 \frac{1}{w_2 b^2} \Bigg\{\uwave{- \; \dot{f}_2 \dot{b}} \; \uwave{\uwave{- \; \omega^2_2 b}} + \frac{w^2_2}{w^2_1} \frac{\omega^2_1}{b^3} \; -\\
    &- \uuline{\; \frac{\dot{b}^2}{b}} \; \uwave{+ \; \dot{f}_2 \dot{b}}\Bigg\} (x^2_2 + y^2_2) \; \uwave{\uline{+ \; l \omega_2 \frac{\psi_1}{b}}} - l \omega_1 \frac{\psi_1}{b} \frac{w_2}{w_1} \frac{1}{b^2} \Bigg] e^{\frac{i}{2} \frac{m}{w_2} \frac{\dot{b}}{b} (x^2_2 + y^2_2)} e^{-i l \int \omega_2 dt_2 + i l \int \omega_1 dt_1}.
\end{align*}
The right-hand side reads
\begin{align*}
    &-\frac{w_2}{2m} \frac{\partial^2 \psi_2}{\partial x^2_2} + \frac{m \omega^2_2 x^2_2}{2 w_2} \psi_2 = \left[-\frac{w_2}{2m} \frac{\partial^2 \psi_1}{\partial x^2_1} \frac{1}{b^3} \; \uline{- \; i \frac{\partial \psi_1}{\partial x_1} \frac{1}{b^2} \frac{\dot{b}}{b} x_2} - \frac{i}{2} \psi_1 \frac{\dot{b}}{b^2} \left\{\uline{\uline{1}} \; \uuline{+ \; i \frac{m}{w_2} \frac{\dot{b}}{b} x^2_2} \right\} \; \uwave{\uwave{+ \; \frac{m \omega^2_2 x^2_2}{2 w_2} \frac{\psi_1}{b}}} \right] \times \\
    &\times e^{\frac{i}{2} \frac{m}{w_2} \frac{\dot{b}}{b} (x^2_2 + y^2_2)} e^{-i l \int \omega_2 dt_2 + i l \int \omega_1 dt_1},
\end{align*}
\begin{align*}
    &- \frac{w_2}{2m} \frac{\partial^2 \psi_2}{\partial y^2_2} + \frac{m \omega^2_2 y^2_2}{2 w_2} \psi_2 = \left[-\frac{w_2}{2m} \frac{\partial^2 \psi_1}{\partial y^2_1} \frac{1}{b^3} \; \uline{-\; i \frac{\partial \psi_1}{\partial y_1} \frac{1}{b^2} \frac{\dot{b}}{b} y_2} - \frac{i}{2} \psi_1 \frac{\dot{b}}{b^2} \left\{\uline{\uline{1}} \; \uuline{+ \; i \frac{m}{w_2} \frac{\dot{b}}{b} y^2_2} \right\} \; \uwave{\uwave{+ \; \frac{m \omega^2_2 y^2_2}{2 w_2} \frac{\psi_1}{b}}} \right] \times \\
    &\times e^{\frac{i}{2} \frac{m}{w_2} \frac{\dot{b}}{b} (x^2_2 + y^2_2)} e^{-il \int \omega_2 dt_2 + il \int \omega_1 dt_1}.
\end{align*}
Next, we assume that the wave function is an eigenstate of the $\hat{L}_z$ operator, i.e. $\psi_1, \psi_2 \sim e^{i l \phi}$. Then action of the $\hat{L}_z = -i \frac{\partial}{\partial \phi}$ for the second system is
\begin{align*}
    \omega_2 \hat{L}_{z_2} \psi_2 &= \omega_2 (-i (i l)) \psi_2 = \uwave{\uline{l \omega_2 \frac{\psi_1}{b} e^{\frac{i}{2} \frac{m}{w_2} \frac{\dot{b}}{b} (x^2_2 + y^2_2)} e^{-il \int \omega_2 dt_2 + il \int \omega_1 dt_1}}},
\end{align*}
and for the first is
\begin{align}
\label{eq:L1}
    \omega_1 \hat{L}_{z_1} \psi_1 &= \omega_1 (-i (i l)) \psi_1 = l \omega_1 \psi_1.
\end{align}
After some simplifications we arrive at
\begin{align*}
    i \frac{\partial \psi_1}{\partial t_1} \frac{w_2}{w_1} \frac{1}{b^3} - \frac{m}{2} \frac{w_2}{w^2_1} \frac{\omega^2_1}{b^3} (x^2_1 + y^2_1) \psi_1 - l \omega_1 \frac{w_2}{w_1} \frac{1}{b^3} \psi_1 = -\frac{w_2}{2 m} \frac{\partial^2 \psi_1}{\partial x^2_1} \frac{1}{b^3} - \frac{w_2}{2 m} \frac{\partial^2 \psi_1}{\partial y^2_1} \frac{1}{b^3}.
\end{align*}
Dividing by $\frac{w_2}{w_1} \frac{1}{b^3}$, with \eqref{eq:L1} we get
\begin{align*}
    i \frac{\partial \psi_1}{\partial t_1} = - \frac{w_1}{2 m} \left(\frac{\partial^2 \psi_1}{\partial x^2_1} + \frac{\partial^2 \psi_1}{\partial y^2_1} \right) + \frac{m \omega^2_1 (x^2_1 + y^2_1)}{2 w_1} \psi_1 + \omega_1 \hat{L}_{z_1} \psi_1.
\end{align*} 
\section{Transformation of the probability current}
We consider a system $\mathcal{H}_2$ where a probability current is given by
\begin{align}
\label{eq:cura}
    \vec{\hat{j_2}} = \frac{1}{m} \left(\sqrt{w_2}\Re{\psi^{*}_2 \vec{\hat{p}_2} \psi_2} - e\vec{A_2} |\psi_2|^2 \right)
\end{align}
First, we consider an action of the operator $\vec{\hat{p}_2} = - i \frac{\partial}{\partial x_2} \vec{e_x} - i \frac{\partial}{\partial y_2} \vec{e_y}$ on $\psi_2$ and then express the result in terms of the system 1 and vectors from $\mathcal{H}_1$:
\begin{align*}
    \vec{\hat{p}_2} \psi_2 = \left[- i \left(\frac{\partial \psi_1}{\partial x_1} \vec{e_x} + \frac{\partial \psi_1}{\partial y_1} \vec{e_y}\right) \frac{1}{b^2} + \frac{m}{w_2} \frac{\dot{b}}{b} \psi_1 (x_1 \vec{e_x} + y_1 \vec{e_y}) \right] e^{\frac{i}{2} \frac{m}{w_2} \frac{\dot{b}}{b} (x^2_2 + y^2_2)}e^{-i l \int \omega_2 dt_2 + i l \int \omega_1 dt_1}
\end{align*}
The first term from \eqref{eq:cura} reads
\begin{align}
\label{eq:cur1}
    \frac{\sqrt{w_2}}{m} \Re{\psi^{*}_2 \vec{\hat{p}_2} \psi_2} = \frac{\sqrt{w_2}}{\sqrt{w_1}} \frac{1}{b^3} \frac{1}{m} \Re{\sqrt{w_1} \psi^{*}_1 \vec{\hat{p}_1} \psi_1} + \frac{\sqrt{w_2}}{\sqrt{w_1}}\frac{1}{m b} |\psi_1|^2 \vec{G},
\end{align}
where $\vec{G} = m \frac{\sqrt{w_1}}{w_2} \frac{\dot{b}}{b} \vec{r_1}$ and $\vec{r_1^{T}} = (x_1,\; y_1,\; 0)$. \\
Next, we consider the transformation of the second term \eqref{eq:cur} under the mapping \eqref{eq:QAT2D_2}. Under symmetric gauge we have
\begin{align*}
    \vec{A_2^{T}} = \left(-\frac{B_2 y_2}{2 \sqrt{w_2}}, \; \frac{B_2 x_2}{2 \sqrt{w_2}}, \; 0\right) = \frac{\sqrt{w_1}}{\sqrt{w_2}} \frac{B_2}{B_1} \left(-\frac{B_1 b y_1}{2 \sqrt{w_1}}, \; \frac{B_1 b x_1}{2 \sqrt{w_1}}, \; 0\right) = \frac{\sqrt{w_1}}{\sqrt{w_2}} \frac{B_2}{B_1} b \vec{A_1^{T}},
\end{align*}
consequently
\begin{align}
\label{eq:cur2a}
    - \frac{e}{m} \vec{A_2} |\psi_2|^2 = - \frac{e}{m} \frac{\sqrt{w_1}}{\sqrt{w_2}} \frac{B_2}{B_1} \frac{1}{b} \vec{A_1} |\psi_1|^2.
\end{align}
We combine \eqref{eq:cur1}, \eqref{eq:cur2a} and arrive at
\begin{align}
    \vec{\hat{j_2}} &= \frac{\sqrt{w_2}}{\sqrt{w_1}} \frac{1}{b^3} \frac{1}{m} \Re{\sqrt{w_1} \psi^{*}_1 \vec{\hat{p}_1} \psi_1} + \frac{\sqrt{w_2}}{\sqrt{w_1}}\frac{1}{m b} |\psi_1|^2 \vec{G} - \frac{e}{m} \frac{\sqrt{w_1}}{\sqrt{w_2}} \frac{B_2}{B_1} \frac{1}{b} \vec{A_1} |\psi_1|^2 = \nonumber \\
    &= \frac{\sqrt{w_2}}{\sqrt{w_1}} \frac{1}{b^3} \vec{\hat{j_1}} + \frac{\sqrt{w_2}}{\sqrt{w_1}} \frac{1}{b^3} \frac{e}{m} \vec{A_1} |\psi_1|^2 + \frac{\sqrt{w_2}}{\sqrt{w_1}}\frac{1}{m b} |\psi_1|^2 \vec{G} - \frac{e}{m} \frac{\sqrt{w_1}}{\sqrt{w_2}} \frac{B_2}{B_1} \frac{1}{b} \vec{A_1} |\psi_1|^2 = \nonumber \\
    &= \frac{\sqrt{w_2}}{\sqrt{w_1}} \frac{1}{b^2}\left\{\frac{\vec{\hat{j_1}}}{b}  +\left[\left(\frac{1}{b} - \frac{B_2}{B_1}\frac{w_1}{w_2} b \right) \frac{e \vec{A_1}}{m}+\frac{b}{m} \vec{G}\right]|\psi_1|^2\right\} .
\end{align}

\end{widetext}

\bibliographystyle{ieeetr}
\bibliography{references}

\end{document}